\documentclass[amsmath,amssymb,prl,aps,mathbbm,superscriptaddress]{revtex4}
\usepackage{amssymb}
\usepackage{graphicx}
\usepackage{times}
\usepackage{subfigure}

\newcommand{\ket}[1]{|#1\rangle}

\usepackage{physics}
\usepackage{graphicx}
\usepackage{amsmath,amssymb,amsthm,dsfont,bm}
\usepackage{color}
\usepackage{soul}
\usepackage{cancel}
\usepackage{braket}
\begin{document}
\title{Digital Quantum Simulation of Linear and Nonlinear Optical Elements}%Attention AE/ME. The following layout issues have not been checked by the English Editing Department and must be carefully verified by the AE/Layout Department: All callout issues, bold usage of callouts, and references to callouts in the text. Correct callout usage in figures. Figure and Table layout issues. Footnote formatting and Glossaries have not been checked. En dash usage for negative values, en dash usage to indicate relationships, en dash usage to indicate bonds (especially in chemistry). The English Editing Department is not responsible for correct italic usage for genes, proteins and technical terminology. This responsibility belongs to the authors. The following are also not checked: spacing between numbers and units of measurement, ratios, en dashes for ranges, date and time formats, punctuation in equation lines, and less than/more than spacing (< >). Finally, capitalization and layout of titles/headings must be properly checked as well as ensuring 'Eq.' and 'Fig.' are properly spelled out, as these are layout issues.

% Authors, for the paper (add full first names)
\author{Carlos Sab\'in}%Please carefully check the accuracy of names and affiliations. Changes will not be possible after proofreading. 
% Authors, for metadata in PDF

% Affiliations / Addresses (Add [1] after \address if there is only one affiliation.)
\affiliation{%
 Instituto de F\'isica Fundamental, CSIC,
Serrano, 113-bis,
28006 Madrid, Spain; csl@iff.csic.es}

% Contact information of the corresponding author
%\corres{Correspondence: casanc05@ucm.es}

% % Current address and/or shared authorship
% \firstnote{Current address: Affiliation 3} 
% \secondnote{These authors contributed equally to this work.}

% Simple summary
%\simplesumm{}

% Abstract (Do not use inserted blank lines, i.e. \\) 
\begin{abstract}
We provide a recipe for the digitalization of linear and nonlinear quantum optics in networks of superconducting qubits. By combining digital techniques with boson-qubit mappings, we address relevant problems that are typically considered in analog simulators, such as the dynamical Casimir effect or molecular force fields, including nonlinearities. In this way, the benefits of digitalization are extended in principle to a new realm of physical problems. We present preliminary examples launched in IBM Q 5 Tenerife. 
\end{abstract}

% Keywords
%\keyword{quantum simulation; quantum computation; quantum optics}

% The fields PACS, MSC, and JEL may be left empty or commented out if not applicable
%\PACS{J0101}
%\MSC{}
%\JEL{}

% If this is an expanded version of a conference paper, please cite it here: enter the full citation of your conference paper, and add $^\S$ in the end of the title of this article.
%\conference{}

%%%%%%%%%%%%%%%%%%%%%%%%%%%%%%%%%%%%%%%%%%
% Only for the journal Data:

%\dataset{DOI number or link to the deposited data set in cases where the data set is published or set to be published separately. If the data set is submitted and will be published as a supplement to this paper in the journal Data, this field will be filled by the editors of the journal. In this case, please make sure to submit the data set as a supplement when entering your manuscript into our manuscript editorial system.}

%\datasetlicense{license under which the data set is made available (CC0, CC-BY, CC-BY-SA, CC-BY-NC, etc.)}

%%%%%%%%%%%%%%%%%%%%%%%%%%%%%%%%%%%%%%%%%%
%\begin{document}\vspace{12pt}

%%%%%%%%%%%%%%%%%%%%%%%%%%%%%%%%%%%%%%%%%%
%% Sections that are not mandatory are listed as such. The section titles given are for Articles. Review papers and other article types have a more flexible structure. 

%% Only for the journal Gels: Please place the Experimental Section after the Conclusions

%%%%%%%%%%%%%%%%%%%%%%%%%%%%%%%%%%%%%%%%%%
\maketitle
After
 decades of both theoretical and experimental efforts, a new generation of technologies is on the brink of delivering the heralded quantum revolution. For instance, large networks of superconducting qubits are close to proving quantum supremacy \cite{preskill,boixo,naturemonty}, opening a new era in 
quantum computing and quantum simulation.

While these promising applications are gaining a great deal of attention, qubits are not the only key players of the quantum world. Indeed, the current scenario has only been possible due to the impressive developments in quantum optical and quantum information setups in the last few decades, which managed to reach the single-atom and single-photon level. Therefore, electromagnetic fields or, more generally, bosonic field modes are also central to modern quantum setups. An~example of the richness offered by the physics of the electromagnetic field is boson sampling \cite{aaaahronson}, namely a~computation of the number statistics of the output photons of a linear optics network, which can in principle be implemented in a quantum optical setup, but is widely believed to be intractable by classical means. Under certain conditions for the number of photons and modes, a boson sampling experiment would also prove quantum supremacy. However, while some small-scale experiments have been realized \cite{experiments1, experiments2, experiments3, experiments4} and a number of promising proposals are available \cite{mbs,dcebs}, a post-classical boson sampler has not yet been implemented in the laboratory. The ingredients needed in a boson sampling architecture are simple linear optics elements, such as beam-splitters and phase-shifters. Going a step further, we can also consider other paradigmatic ingredients such as two-mode squeezers or Kerr nonlinearities. Indeed, the combination of controllable two-mode squeezers and beam-splitter interactions between two bosonic modes is the basis of quantum information processing and quantum computation with continuous variables \cite{gao,zhanggirvin}, such as the realization of Fredkin and exponential SWAP%define if appropriate
 gates, which have recently been implemented experimentally with fidelities ranging from $60\%$ to 90\%~\cite{schoelkopfgirvin}. Kerr nonlinearities can also find interesting applications, such as quantum metrology \cite{napolitano}, whose simulation in a quantum computer might thus be of interest.

In this work, we aim to bring linear and nonlinear optics to the realm of cutting-edge qubit-based quantum computers and quantum simulators. To this end, we consider digital quantum simulation techniques \cite{lloyd}, which combined with a boson-qubit mapping \cite{losalamos,sommathesis,anal}, allow us to encode generic multimode bosonic interactions into a sequence of single-qubit and two-qubit gates. Applications include the digital quantum simulation of continuous-variable quantum information processing and computing \cite{electrophonon,fermiboson,arxiv1,arxiv2}, as well as phenomena that are typically considered in analog quantum simulators, from the dynamical Casimir effect \cite{dce,entanglementchalmers,discord,steering,coherence,felicetti,andres,DCE-BECs,DCE-ions} to molecular force fields \cite{thesisjoonsuk,elblueprint,anal,review}. While~the digital quantum simulation of a post-classical boson sampler seems completely out of reach, we~discuss the digitalization of all the necessary ingredients for boson sampling, which would enable the digitalization of a few-mode setup. 

In this way, a new manifold of multidisciplinary problems of interest---ranging from quantum chemistry to relativistic quantum field theory---could benefit in principle from the advantages of digitalization, which have already proven successful in many applications, from universal trapped-ion simulators \cite{universal} to fermionic models \cite{fermionic} and adiabatic quantum computing \cite{digitizedadiabatic} with superconducting circuits. On the other hand, the existing and upcoming quantum computers based on large arrays of superconducting qubits \cite{geller} would also benefit in principle from a new variety of problems of interest. We present some preliminary experimental examples obtained with IBM Q 5 Tenerife. 

Let us start with a description of the technical tools that we shall use throughout this work for the digitalization of bosonic Hamiltonians.

%\section{Methods}

\section{Digitalization of Bosonic Hamiltonians}%Section number has been changed, please confirm.

The aim of this work is to provide a recipe for the quantum simulation of bosonic Hamiltonians of interest with networks of qubits. The first step is to encode the bosonic Hamiltonian into the qubit network by means of a suitable boson-qubit mapping.

\subsection{Mapping Bosons to Qubits}

As shown in \cite{losalamos, sommathesis}, it is possible to map N bosonic modes containing a maximum number of $N_p$ excitations each to $N(N_p+1)$ qubits. 
For each bosonic mode, we are able to associate an $(N_P+1)$-qubit quantum state to each Fock state. If we consider the $j^{\text{th}}$ mode, we have:
\begin{eqnarray}
\label{bosonmap}
|0\rangle_j &\leftrightarrow& |0_0 1_1 1_2 \cdots 
1_{N_P} \rangle_j \nonumber \\
|1\rangle_j &\leftrightarrow& |1_0 0_1 1_2 \cdots 
1_{N_P} \rangle_j \nonumber \\
|2\rangle_j &\leftrightarrow& |1_0 1_1 0_2 \cdots 
1_{N_P} \rangle_j \\
\vdots && \vdots \nonumber \\
|N_P\rangle_j &\leftrightarrow& |1_0 1_1 1_2
\cdots 0_{N_P} \rangle_j \nonumber
\end{eqnarray} 
where $\ket{n}_j$ denotes a quantum state with $n$ bosons in the $j^{\text{th}}$ mode. 
Notice that the state $\ket{n}_j$ is simulated by a state where out of the $N_P+1$ qubits that are associated with the $j^{\text{th}}$ mode, the only one that is in the state $\ket{0}$ is the $n^{\text{th}}$ qubit.
 
The bosonic creation operator maps to:
\begin{equation}
\label{bosonmap2}
b^{\dagger}_j \rightarrow b^{\dagger}_j =
\sum \limits_{n=0}^{N_P-1} \sqrt{n+1} \ \sigma_-^{n,j}
\sigma_+^{n+1,j} ,
\end{equation}
where a pair $(n,j)$ refers to the $n^{\text{th}}$ qubit in the chain of qubits
representing the $j^{\text{th}}$ bosonic mode. The~Pauli creation and annihilation
operators are given by: 
\begin{equation}\label{eq:qubitpm}
\sigma_\pm^k=1/2 (\sigma_x^k\pm i\sigma_y^k),
\end{equation}
in terms of the Pauli matrices $\sigma_x$ and $\sigma_y$. From Equation (\ref{eq:qubitpm}), it is straightforward to obtain as well the annihilation operator and then all the combinations that appear in bosonic Hamiltonians of interest; we will see relevant examples in the next sections.

Once we have encoded the bosonic Hamiltonian into a suitable qubit-network Hamiltonian and before we enter into the details of examples and applications, let us recall the basic notion of digital quantum simulations, namely the Suzuki--Trotter approximation.

\subsection{Trotter--Suzuki%or Suzuki--Trotter? please check the convention throughout
 Decomposition}

The idea of the Trotter--Suzuki decomposition (see for instance \cite{lloyd}) is to decompose an involved non-local Hamiltonian dynamics into a sequence of experimentally amenable local gates. To this end:
\begin{itemize}
\item The Hamiltonian is decomposed into a number $m$ of terms:
\begin{equation}
H=\sum_{j=1}^m H_j
\end{equation}
\item Time is discretized, namely divided into $s$ steps of duration $t_s$:
\begin{equation}
s=t/t_s.
\end{equation}
\item The exponential of a sum of operators is approximated by the exponential of a product of operators. The approximation is exact only in the case that all the Hamiltonian terms commute. Otherwise, it neglects all the commutators in the Baker--Campbell--Hausdorff formula.
\end{itemize}

Putting it all together, the dynamics is approximated by:
\begin{equation}
e^{iHt}\simeq (e^{iH_1t_s}\dots e^{iH_m t_s})^s,
\end{equation}
that is $s$ repetitions of the sequence of Hamiltonian terms given by $H_j$. The error in this approximation can be bounded and controlled \cite{lloyd}, being suppressed by a sufficiently large number of repetitions. However, the number of repetitions also increases the number of gates, which gives rise to larger experimental errors. Therefore, in the cases where the Suzuki--Trotter is exact---namely the different terms of the Hamiltonian commute---there is no need for repetitions.

The Suzuki--Trotter approach allows us to simulate any Hamiltonian dynamics as a sequence of unitary operations. The next step is now to decompose any unitary as a sequence of simple quantum gates, typically chosen from a desired set of universal single-qubit and two-qubit quantum gates. 

\subsection{Gate Decomposition}

In general, if we find an unitary operation $U$ such that:
\begin{equation}\label{eq:gatedec1}
H=U^{\dagger}H_0 U,
\end{equation}
then we can write the dynamics governed by the Hamiltonian $H$ as:
\begin{equation}\label{eq:gatedec2}
e^{i H t}=U^{\dagger}e^{i H_0 t} U.
\end{equation}
The idea is then to relate $H$ with a simple single-qubit $H_0$ via an unitary $U$. Then, this $U$ can be decomposed into a sequence of single-qubit and two-qubit gates of interest, by using well-known techniques \cite{sommalallama}. We will analyze some examples of gate decompositions below.

%\section{Results}

\section{Examples}

We will apply the techniques of the previous section to several families of interesting bosonic Hamiltonians.

\subsection{Boson Sampling and Boson Sampling Hamiltonian}

Boson sampling consists of the sampling of the output photon-number statistics of a linear-optics network operating over a number $M$ of photonic modes, which are initialized in a certain Fock state containing $n$ photons. It has been shown \cite{aaaahronson} that in the regime where $M\simeq n^2$ and $n\simeq 20-50$ \cite{noimminent}, this problem is computationally hard and should be intractable by classical devices. While this feature turns a boson sampler into a compelling architecture for quantum supremacy, the experiments \cite{experiments1,experiments2,experiments3,experiments4} have not yet reached the aforementioned post-classical regime.

Recently, a Hamiltonian formulation of boson sampling was introduced \cite{spinsampling}: 
\begin{equation}
H_{BS} = \sum_{i,j=1}^M( b_j^\dagger R_{ji} a_i + \mathrm{H.c.})
+ \sum_{j=1}^M \omega(b^\dagger_jb_j + a^\dagger_ja_j),
\label{eq:harmonic}
\end{equation}
where $R$ is the matrix representation of a random unitary transformation and $a$ and $b$ are the input and output modes, respectively. Evolution with this Hamiltonian transforms an initial Fock state:
\begin{equation}
\ket{\phi(0)} = a_1^\dagger \cdots a_N^\dagger\ket{\mathrm{vac}},
\label{eq:bs-input}
\end{equation}
into the $N$ boson sampling superposition:
\begin{equation}
\ket{\phi(\pi/2)} = (-i)^N\prod_{i=1}^N \sum_j R^*_{ji} b_j^\dagger\ket{\mathrm{vac}},
\label{eq:bs-state}
\end{equation}
with a photon distribution given by the permanents $
|\gamma_{\mathbf{n}}|^2 = |\braket{\mathrm{vac}|b_1^{\dagger n_1}\cdots b_M^{\dagger n_M}|\phi(\pi/2)}|^2$, $n_i\in \{0,1\}$.
Therefore, our task would be to digitalize the Hamiltonian in Equation (\ref{eq:harmonic}). In this case, we do not need to use the Suzuki--Trotter decomposition since we can directly leverage the Reck decomposition, which~is standard in the boson sampling literature and entails that we can decompose the unitary evolution governed by $H_{BS}$ into a mesh of $M (M-1)/2$ beam-splitters and appropriate phase-shifters. Each~beam-splitter unitary would be given by:
\begin{equation}
U_{ij}=e^{i\varepsilon_{ij}b^{\dagger}_ja_i+h.c.}.\label{eq:beams}
\end{equation}
Since they are the building blocks of the boson sampling simulation, it is worth analyzing in detail the digitalization of these unitary beam-splitting interactions. Moreover, beam-splitter interactions are also crucial for applications in continuous-variables quantum information processing and quantum computing \cite{gao,zhanggirvin,schoelkopfgirvin}.

\subsection{Beam-Splitters}

The initial state of boson sampling contains a maximum number of one photon per mode; therefore, the first beam-splitter of the mesh would only require 2 modes $\times$ 2 qubits per mode $=$ 4 qubits. Then,~it is interesting to consider in detail this case.

We will label the two modes of interest as $+$ and $-$, and thus, a beam-splitter unitary:
\begin{equation}\label{eq:bs}
U_{+-}=e^{i\varepsilon_{+-}b^{\dagger}_+a_-+h.c.}.
\end{equation}
We will need two qubits $0+$ and $1+$ for mode $+$ and similarly for mode $-$. Particularizing the boson-qubit operator mapping in Equation (\ref{bosonmap2}) for $N_P=1$, we can write:
\begin{eqnarray}
b_+^\dagger b_-+b_+b_-^\dagger&=&\sigma_-^{(0+)}\sigma_+^{(1+)}\sigma_+^{(0-)}\sigma_-^{(1+)}+\\\nonumber & &\sigma_+^{(0+)}\sigma_-^{(1+)}\sigma_-^{(0-)}\sigma_+^{(1-)}.
\end{eqnarray}
Now, let us relax the notation:
\begin{equation}\label{eq:relax}
(0+)=(1),\,\,\,\, (1+)=(2),\,\,\,\, (0-)=(3,)\,\,\,\, (1-)=(4).
\end{equation}
By using Equation (\ref{eq:qubitpm}), we find:
\begin{eqnarray}\label{eq:mappedbs}
b_+^\dagger b_-+b_+b_-^\dagger &=&\frac{1}{8}(\sigma_x^{(1)}\sigma_x^{(2)}\sigma_x^{(3)}\sigma_x^{(4)}-\sigma_x^{(1)}\sigma_y^{(2)}\sigma_y^{(3)}\sigma_x^{(4)}+\nonumber\\ & &\sigma_x^{(1)}\sigma_y^{(2)}\sigma_x^{(3)}\sigma_y^{(4)}+\sigma_x^{(1)}\sigma_x^{(2)}\sigma_y^{(3)}\sigma_y^{(4)}+\nonumber \\& &\sigma_y^{(1)}\sigma_y^{(2)}\sigma_x^{(3)}\sigma_x^{(4)}+\sigma_y^{(1)}\sigma_x^{(2)}\sigma_y^{(3)}\sigma_x^{(4)}-\nonumber\\ &&\sigma_y^{(1)}\sigma_x^{(2)}\sigma_x^{(3)}\sigma_y^{(4)}+\sigma_y^{(1)}\sigma_y^{(2)}\sigma_y^{(3)}\sigma_y^{(4)}).
\end{eqnarray}
Although it might seem that the different terms in Equation (\ref{eq:mappedbs}) do not commute, we have checked that they do. Consider for instance the first two terms and notice that, using the properties of the commutators and the fact that the operators acting on different qubits obviously commute, we have:
\begin{eqnarray}\label{eq:commutators}
&&[\sigma_x^{(1)}\sigma_x^{(2)}\sigma_x^{(3)}\sigma_x^{(4)}, \sigma_x^{(1)}\sigma_y^{(2)}\sigma_y^{(3)}\sigma_x^{(4)}]=[\sigma_x^{(1)}\sigma_x^{(2)},\sigma_x^{(1)}\sigma_y^{(2)}]\nonumber\\& &\sigma_x^{(3)}\sigma_x^{(4)}\sigma_y^{(3)}\sigma_x^{(4)}+ \sigma_x^{(1)}\sigma_y^{(2)}\sigma_x^{(1)}\sigma_x^{(2)}[\sigma_x^{(3)}\sigma_x^{(4)},\sigma_y^{(3)}\sigma_x^{(4)}]
\end{eqnarray}
Then, similarly:
\begin{eqnarray}
[\sigma_x^{(1)}\sigma_x^{(2)},\sigma_x^{(1)}\sigma_y^{(2)}]=(\sigma_x^{(1)})^2
[\sigma_x^{(2)},\sigma_y^{(2)}] \end{eqnarray}
\begin{eqnarray}
 [\sigma_x^{(3)}\sigma_x^{(4)},\sigma_y^{(3)}\sigma_x^{(4)}] =[\sigma_x^{(3)},\sigma_y^{(3)}](\sigma_x^{(4)})^2 \end{eqnarray}
Putting everything together and using the properties of Pauli matrices
$\sigma_i^{(k)}\sigma_j^{(k)}=i\epsilon_{ijk}\sigma_k^{(k)}$ ($\epsilon_{ijk}$ being the Levi--Civita tensor), the two contributions in Equation (\ref{eq:commutators}) cancel out and similarly with all the terms in Equation (\ref{eq:mappedbs}).

\begin{figure}[h]\centering
\includegraphics[width=\textwidth]{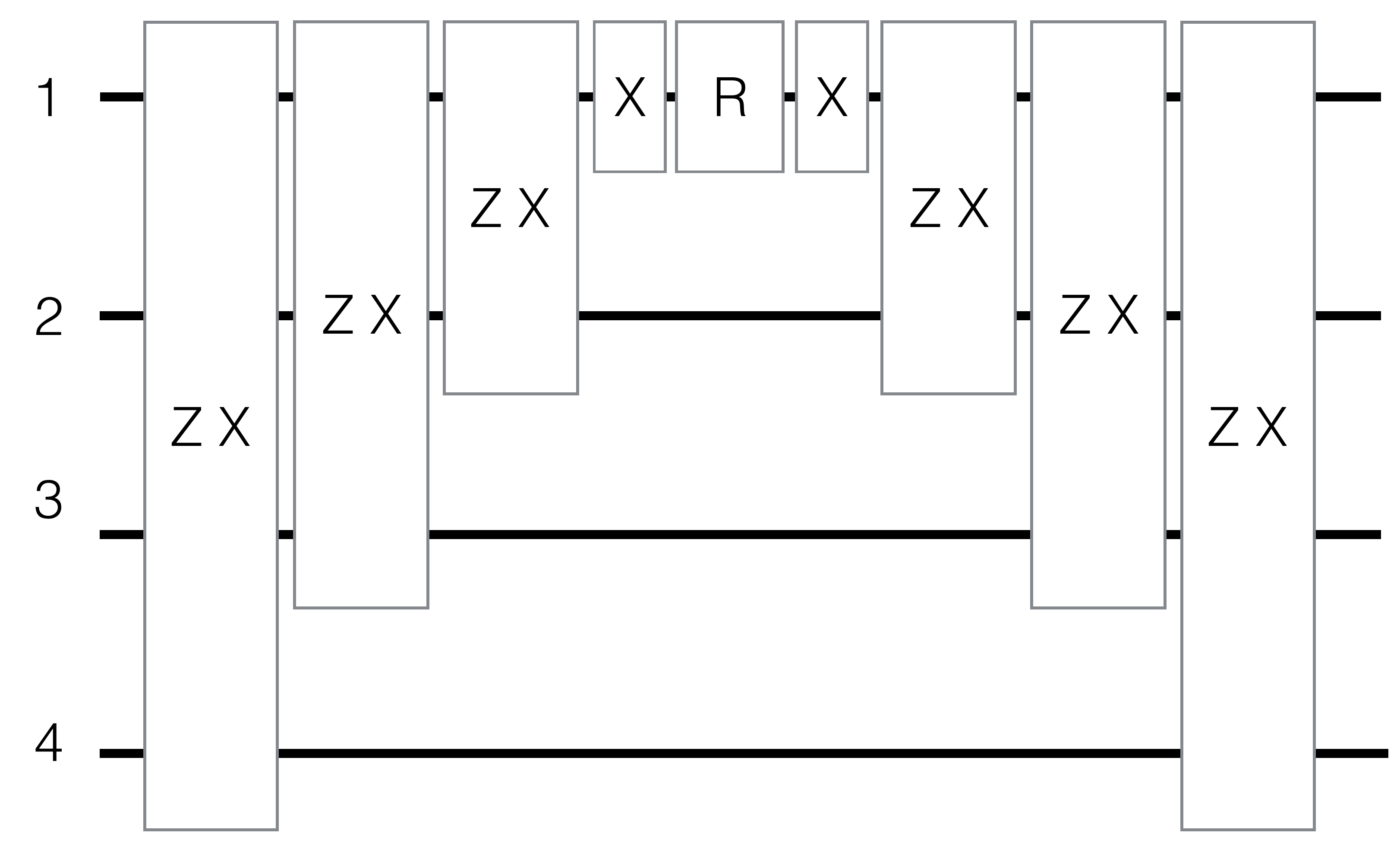} 
\caption{Sequence of gates that implement the beam-splitting interaction term $U_{+-}^{(1)}$ in Equation (\ref{eq:upm1}). $R=e^{-i\frac{\varepsilon_{+-}}{8}\sigma_z^{(1)}}$, $ZX=e^{i\frac{\pi}{4}\sigma_z^{(i)}\sigma_x^{(j)}}$, where i%should it be italics? please check the conventions for the math notations throughout
 and j stand for the qubits among which the two-qubit gate operates, and $X=e^{i\frac{\pi}{4}\sigma_x^{(1)}}$. The rest of the terms $U_{+-}^{(i)}$ possess a similar gate decomposition.}
 \label{Fig1}
 \end{figure}

Since everything commutes, we have: 
\begin{equation}
U_{+-}=\prod_{i=1}^8 U_{+-}^{(i)},
\end{equation}
where the $ U_{+-}^{(i)}$'s are given by Equation (\ref{eq:mappedbs}). For instance:
\begin{equation}\label{eq:mappedbsu1}
U_{+-}^{(1)}=e^{i\frac{\varepsilon_{+-}}{8}\sigma_x^{(1)}\sigma_x^{(2)}\sigma_x^{(3)}\sigma_x^{(4)}}.
\end{equation}
Then, we can perform a separate gate decomposition for each $U_{+-}^{(i)}$. For instance, for the first term, we can use:
\begin{eqnarray}
e^{-i\frac{\pi}{4}\sigma_x^{(1)}}(-\sigma_z^{(1)})e^{i\frac{\pi}{4}\sigma_x^{(1)}}&=&\sigma_y^{(1)}\\\nonumber
e^{-i\frac{\pi}{4}\sigma_z^{(1)}\sigma_x^{(2)}}\sigma_y^{(1)}e^{i\frac{\pi}{4}\sigma_z^{(1)}\sigma_x^{(2)}}&=&-\sigma_x^{(1)}\sigma_x^{(2)}\\\nonumber
e^{-i\frac{\pi}{4}\sigma_z^{(1)}\sigma_x^{(3)}}(-\sigma_x^{(1)}\sigma_x^{(2)})e^{i\frac{\pi}{4}\sigma_z^{(1)}\sigma_x^{(3)}}&=&-\sigma_y^{(1)}\sigma_x^{(2)}\sigma_x^{(3)}\\\nonumber
e^{-i\frac{\pi}{4}\sigma_z^{(1)}\sigma_x^{(4)}}(-\sigma_y^{(1)}\sigma_x^{(2)}\sigma_x^{(3)})e^{i\frac{\pi}{4}\sigma_z^{(1)}\sigma_x^{(4)}}&=&\sigma_x^{(1)}\sigma_x^{(2)}\sigma_x^{(3)}\sigma_x^{(4)}.
\end{eqnarray}
Then, by using Equations (\ref{eq:gatedec1}), (\ref{eq:gatedec2}), and (\ref{eq:mappedbsu1}), we find:
\begin{equation}
U_{+-}^{(1)}=U^\dagger e^{-i\frac{\varepsilon_{+-}}{8}\sigma_z^{(1)}}U,\label{eq:upm1}
\end{equation}
where:
\begin{equation}
U= e^{i\frac{\pi}{4}\sigma_x^{(1)}}e^{i\frac{\pi}{4}\sigma_z^{(1)}\sigma_x^{(2)}}e^{i\frac{\pi}{4}\sigma_z^{(1)}\sigma_x^{(3)}}e^{i\frac{\pi}{4}\sigma_z^{(1)}\sigma_x^{(4)}}.\label{eq:u1}
\end{equation}
See this gate decomposition in Figure \ref{Fig1}. Note that similar decompositions can be obtained for the rest of the $ U_{+-}^{(i)}$'s by adding at the end of the string the number of $e^{i\pi/4\sigma_z^{(i)}}$ necessary to rotate some of the $\sigma_x$ to $\sigma_y$; although there might be more efficient decompositions for each particular case. With this procedure, we will have a total number of 24 single-qubit rotations and 24 two-qubit $ZX$-gates.

The relevant initial states would be for instance $|0\rangle_+\otimes|1\rangle_-$ and $|1\rangle_+\otimes|0\rangle_-$, which via Equations (\ref{bosonmap}) and (\ref{eq:relax}),
are mapped to $|0110\rangle$ and $|1001\rangle$, respectively, which can be obtained by spin-flipping a~pair of qubits out of the four-qubit ground state.

\begin{figure}[h]\centering
\includegraphics[width=0.98\textwidth]{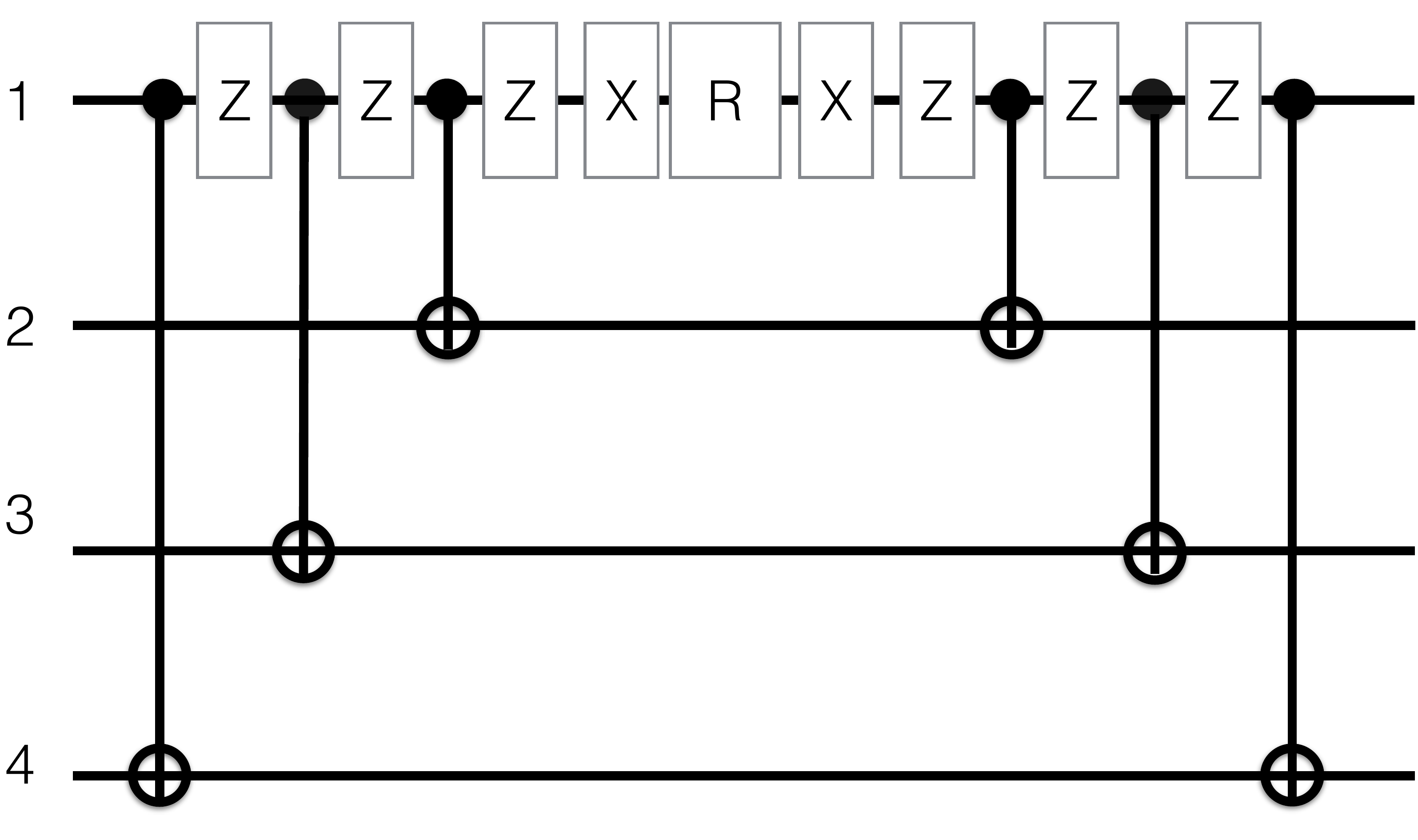} 
\caption{Sequence of gates that implement the beam-splitting interaction term $U_{+-}^{(1)}$ in Equation (\ref{eq:upm1}) after transforming the ZX%define if appropriate
 gates into CNOT%define if appropriate
 gates. $R=e^{-i\frac{\varepsilon_{+-}}{8}\sigma_z^{(1)}}$, $Z=e^{i\frac{\pi}{4}\sigma_z^{(1)}}$, and $X=e^{i\frac{\pi}{4}\sigma_x^{(1)}}$.}
 \label{fig1-bis}
 \end{figure}

Putting all the above together, a single two-mode beam-splitter with one photon per mode can be simulated in a four-qubit quantum simulator. Indeed, we launched the experiment in Figure \ref{Fig1} for the initial state $|1\rangle_+\otimes|0\rangle_-$ in IBM Q 5 Tenerife by means of the qiskit tools. An extra step is needed for that, which is the conversion of the ZX%define if appropriate
 gates into the CNOT%define if appropriate
 gates available in the IBM architecture. This can be done by using:
\begin{equation}
e^{i\frac{\pi}{4}\sigma_z^{(1)}\sigma_x^{(2)}}= e^{i\frac{\pi}{4}\sigma_x^{(2)}}e^{i\frac{\pi}{4}\sigma_z^{(1)}}e^{-i\frac{\pi}{4}} CNOT^{(1-2)} \label{eq:cnot}
\end{equation}
(and similarly with $1-3$ and $1-4$%please check the convention throughout
). Inserting Equation (\ref{eq:cnot}) into Equations (\ref{eq:upm1}) and (\ref{eq:u1}) and making straightforward simplifications, we get the circuit in Figure \ref{fig1-bis}.

In particular, we choose $\frac{\varepsilon_{+-}}{8}=\pi$, so the dynamics should completely transform the initial state into $|0\rangle_+\otimes|1\rangle_-$, namely $|0110\rangle$. We~check the state of the first qubit, which is the one involved in more quantum gates (after running the circuit). We get the right state in 1717 out of 2048 shots, namely 84\%. Running the whole beam-splitter---that is, the eight terms of Equation (\ref{eq:mappedbs})---with the same parameters, the state of the first qubit should be one, and we get the right state 1158 shots out of 2048 (57\%). Note~that a change in the initial state is not expected to modify the fidelity significantly, since the number of required gates is not significantly~affected.

\subsection{Sequence of Beam Splitters}

If we allow more modes and more photons per mode in the simulation, the number of required qubits increases dramatically. However, since a beam-splitter only acts upon two modes and the transition from $|n\rangle_j$ to $|n\pm1\rangle_j$ only flips two qubits, the action of a beam-splitter onto a definite Fock state $|n\rangle_j\otimes|m\rangle_k$ with $n$ photons in mode $j$ and $m$ photons in mode $k$ can still be implemented in a~few-qubit quantum simulator. However, this is not enough for a boson-sampling architecture. After~a~number of beam-splitting steps, the state of the modes would not be a definite Fock state, but~a~non-trivial combination of them. Therefore, a large number of qubits would be involved at each step, a~signature of the computational complexity of boson sampling.

In the post-classical regime, $N_P\simeq 20 - 50$ and the number of modes $N_P^2$. Therefore, the full simulation would require $10^4-10^5$%please check the convention throughout
 qubits, a number that seems completely out of reach.

\subsection{Two-Mode Squeezing}
 
In a Gaussian boson sampler \cite{Lund2014}, the initial state is Gaussian, as opposed to the initial Fock state of the standard boson sampling. A particular architecture is a setup in which half of the output of $M$ two-mode squeezers is input into a linear network of $M$ optical modes, while the other half is sent directly to single photon detectors. Then, $n$ single photons are detected in the latter half. As~shown in~\cite{Lund2014}, this device is able to solve a randomized version of boson sampling known as scattershot boson sampling, which possesses a similar computational complexity as the original problem \cite{aaaahronson} and~therefore is widely believed to be out of reach classically.

\begin{figure}[h]\centering
\includegraphics[width=0.98\textwidth]{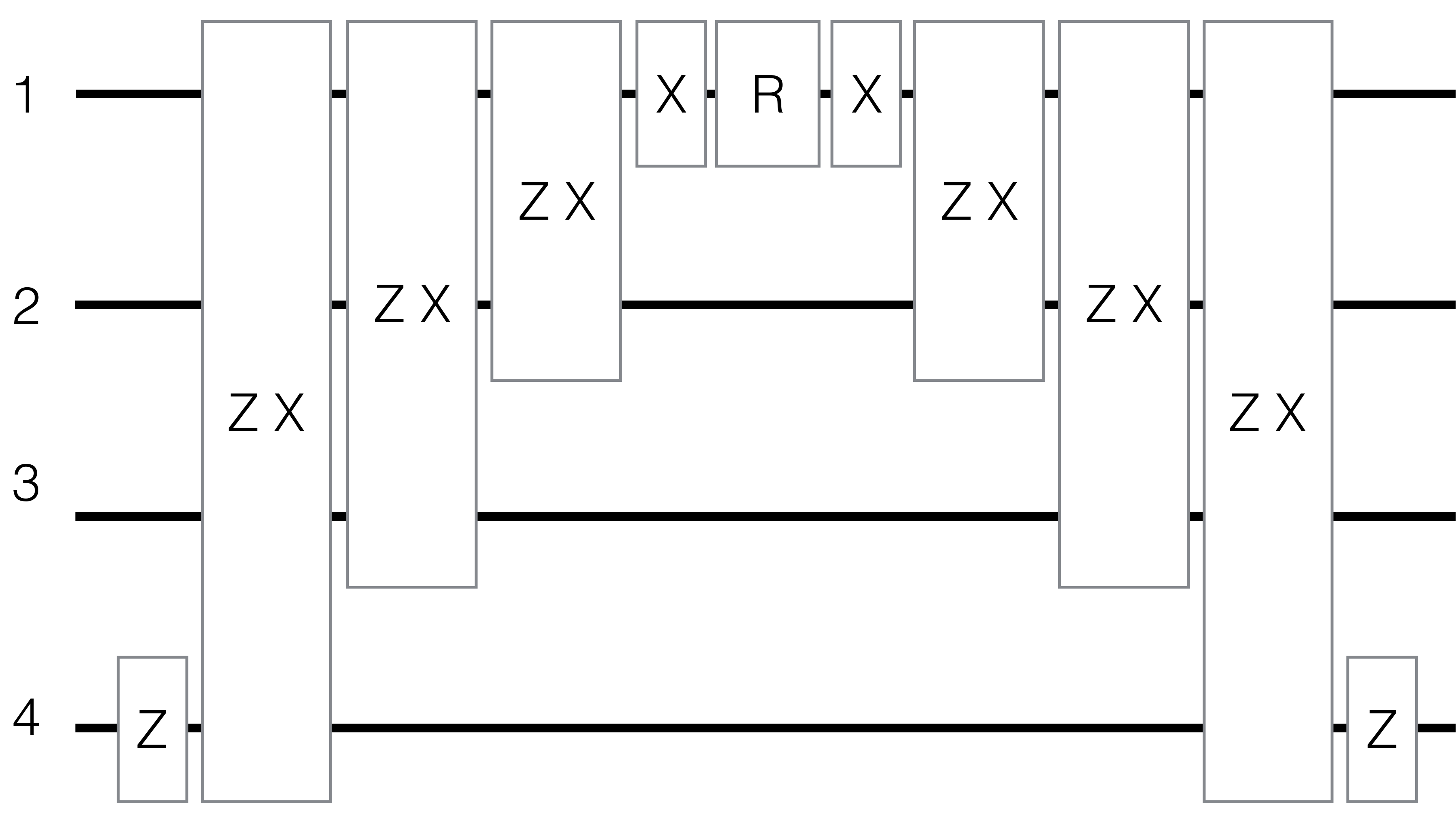} 
\caption{Sequence of gates that implement the two-mode squeezer term $\sigma_x^{(1)}\sigma_x^{(2)}\sigma_x^{(3)}\sigma_y^{(4)}$. \mbox{$R=e^{-i\frac{\beta_{+-}}{8}\sigma_z^{(1)}}$, $ZX=e^{i\frac{\pi}{4}\sigma_z^{(i)}\sigma_x^{(j)}}$}, where i and j stand for the qubits among which the two-qubit gate operates, and $X=e^{i\frac{\pi}{4}\sigma_x^{(1)}}$, $Z=e^{i\frac{\pi}{4}\sigma_z^{(4)}}$. }
 \label{Fig2}
 \end{figure}

To simulate this, we only need to add to our recipe the ability of simulating an initial set of two-mode squeezed states. For each of two modes, we only need to initialize them in the vacuum state $|0101\rangle$ and then apply the unitary:
\begin{equation}\label{eq:tms}
U_{+-}=e^{i\beta_{+-}a^{\dagger}_+a^{\dagger}_-+h.c.}.
\end{equation}
Restricting ourselves to moderate values of the squeezing parameter $\beta$, the probability of generating more than one photon pair would be low, and then, we would be able to assume $N_P=1$. In particular, the average number of photons per mode would be given by $\sinh^2{\beta}$ \cite{carltms}, so for $\beta=0.5$, we have an~average number of photons of 0.3, and the approximation is still safe. Note that the approximation of a~maximum of one photon per mode is also assumed in scattershot boson sampling \cite{Lund2014}. Therefore, again~only four qubits per pair of modes would suffice. For each two-mode squeezer, similar techniques as in the case of the beam-splitter can be applied, with similar results in terms of computational complexity. Indeed, for each two-mode squeezer, we get a similar expression to Equation (\ref{eq:mappedbs}), but now instead of having products of even numbers of $\sigma_x$ and $\sigma_y$, we have products of odd numbers, such~as $\sigma_x^{(1)}\sigma_x^{(2)}\sigma_x^{(3)}\sigma_y^{(4)}$, which give rise to similar terms as in Equations (\ref{eq:upm1}) and (\ref{eq:u1}). Moreover, they also commute with each other. In particular, as in the beam-splitter terms, they can be obtained by adding as many $e^{i\pi/4\sigma_z^{(i)}}$ as the required $\sigma_y^{(i)}$ to $U$ in Equation (\ref{eq:u1}). See Figure \ref{Fig2} for the gate decomposition corresponding to $\sigma_x^{(1)}\sigma_x^{(2)}\sigma_x^{(3)}\sigma_y^{(4)}$.
 
Note that larger values of squeezing can already be achieved in other systems, such as cold atom setups \cite{squeez1,squeez2,squeez3,squeez4}. In our case, we would need to increase the number of excitations allowed and therefore the number of qubits. Note also that we are not considering three-mode or multimode squeezing Hamiltonians, which can be achieved for instance in cold atoms \cite{squeez3} and superconducting circuits \cite{threePRX} and which would give rise to completely different multimode states \cite{threePRX,andres2}.
\subsection{Bogoliubov Transformations}

Combining the techniques for digitalizing beam-splitters and two-mode squeezers, we can consider the simulation of arbitrary Bogoliubov transformations, which transform any annihilation operator $a_i$~to:
\begin{equation}\label{eq:bogobos}
\widehat{a}_i=\sum_{j}\alpha_{ji}a_i+\beta_{ji}a^{\dagger}_i.
\end{equation}
From a Hamiltonian viewpoint, the transformation in Equation (\ref{eq:bogobos}) amounts to a set of two-mode beam-splitters implemented by the $\alpha_{ji}$ coefficients and two-mode squeezers implemented by the $\beta_{ji}$ coefficients. Therefore, in principle, any Bogoliubov transformation could be simulated using the techniques developed in this paper. 

Bogoliubov transformations are ubiquitous in quantum field theory, especially in relativistic applications such as the dynamical Casimir effect, the Unruh effect, and Hawking radiation. An~immediate application would be a digital quantum simulation of the dynamical Casimir effect, namely the generation of photons out of the vacuum of a quantum field by means of the relativistic motion of a mirror, which generates two-mode squeezing. For the simplest case of two modes and moderate values of the squeezing---which is proportional to the velocity of the moving mirror---four qubits would be enough, as in the case of the previous subsection. Considering higher values of the squeezing and thus higher values of $N_P$ would allow considering larger velocities, presumably overcoming the limitations in analog quantum simulators \cite{dce}. Note that we only need to go to $N_p=2$, since~the maximum value of squeezing achieved in experiments is 0.46 \cite{dce}, which can be safely simulated with $N_p=1$, as discussed above.

\subsection{Quantum Information Processing and Quantum Computing Gates}

There is growing effort, both theoretically and experimentally \cite{gao,zhanggirvin,schoelkopfgirvin}, in continuous-variable quantum information and quantum computation with multiphoton states in superconducting microwave cavities. As explained for instance in \cite{gao}, a key step is the ability of generating a~controllable set of bilinear interactions, namely an interaction Hamiltonian of the form:
\begin{eqnarray}
H_{int}&=& g_{BS} (t) a^{\dagger}b\,+\,g_{BS} (t) ab^{\dagger}\,\nonumber\\ &+&\,g_{TMS} (t) a^{\dagger}b^{\dagger}\,+\,g_{TMS} (t) ab,\label{eq:bilinear}
\end{eqnarray}
which is a combination of beam-splitting and two-mode squeezing terms. This interaction Hamiltonian can be simulated with our techniques. In \cite{schoelkopfgirvin}, a Hamiltonian like Equation (\ref{eq:bilinear}) with only beam-splitting terms was used to generate experimentally several gates, with a fidelity ranging from 60 to 90\%. Therefore, it could be of interest to compare these results with a digital quantum simulation, with~could reach similar fidelities and benefit from quantum error correction techniques.

\subsection{Molecular Force Fields}

Molecular transitions can be described by using a set of bosonic modes immersed in a force field, which can be modeled by a non-harmonic oscillator \cite{thesisjoonsuk, elblueprint}. For each mode, we will have:
\begin{equation}\label{eq:molecules}
H_j= \omega_j n_j + \chi_j n_j^2,
\end{equation}
where $n_j=a_j^{\dagger}a_j$ is the number operator of the j$^{\text{th}}$ mode and the parameter $\chi_j$ accounts for the anharmonicity of the Hamiltonian. The mapped number operator is \cite{losalamos,sommathesis}:
\begin{equation}
\label{bosonmap2d}
n_j =\sum \limits _{n=0}^{N_P} n \
\frac{\sigma_z^{n,j}+1}{2} ,
\end{equation}
and then:
\begin{equation}
\label{bosonmap2d2}
n_j^2 =\sum \limits _{n=0}^{N_P} \sum \limits _{m=0}^{N_P}n \
\frac{\sigma_z^{n,j}+1}{2} m \
\frac{\sigma_z^{m,j}+1}{2}.
\end{equation}
Typically, it is enough to consider three or four excitations per mode. Considering for instance $N_P=4$, we~would need five qubits per mode. Then, expanding explicitly Equations (\ref{bosonmap2d}) and (\ref{bosonmap2d2}), we obtain:

\begin{equation}
\label{bosonmap2d4}
n_j =5+\sum\limits_{n=0}^{4}\frac{n}{2}\sigma_z^{n,j} ,
\end{equation}
and: 
\begin{equation}\label{bosonmap2d42}
n_j^2 =\frac{65}{2}+
\sum \limits _{n=0}^{4} 
5\,n\,\sigma_z^{n,j}+\sum \limits _{n=0}^{4}\sum \limits _{m\neq n}\frac{n\,m}{2} \sigma_z^{n,j}\sigma_z^{m,j}.
\end{equation}
Note that all the terms in Equations (\ref{bosonmap2d4}) and (\ref{bosonmap2d42}) commute. Therefore, the Suzuki--Trotter decomposition is exact. Moreover, each term is already given by either single-qubit or two-qubit gates. In particular, we~have four single-qubit rotations for the simulation of $n_j$ and additionally four single-qubit and six two-qubit gates for the simulation of $n_j^2$. 

\begin{figure}[h]\centering
\includegraphics[width=0.98\textwidth]{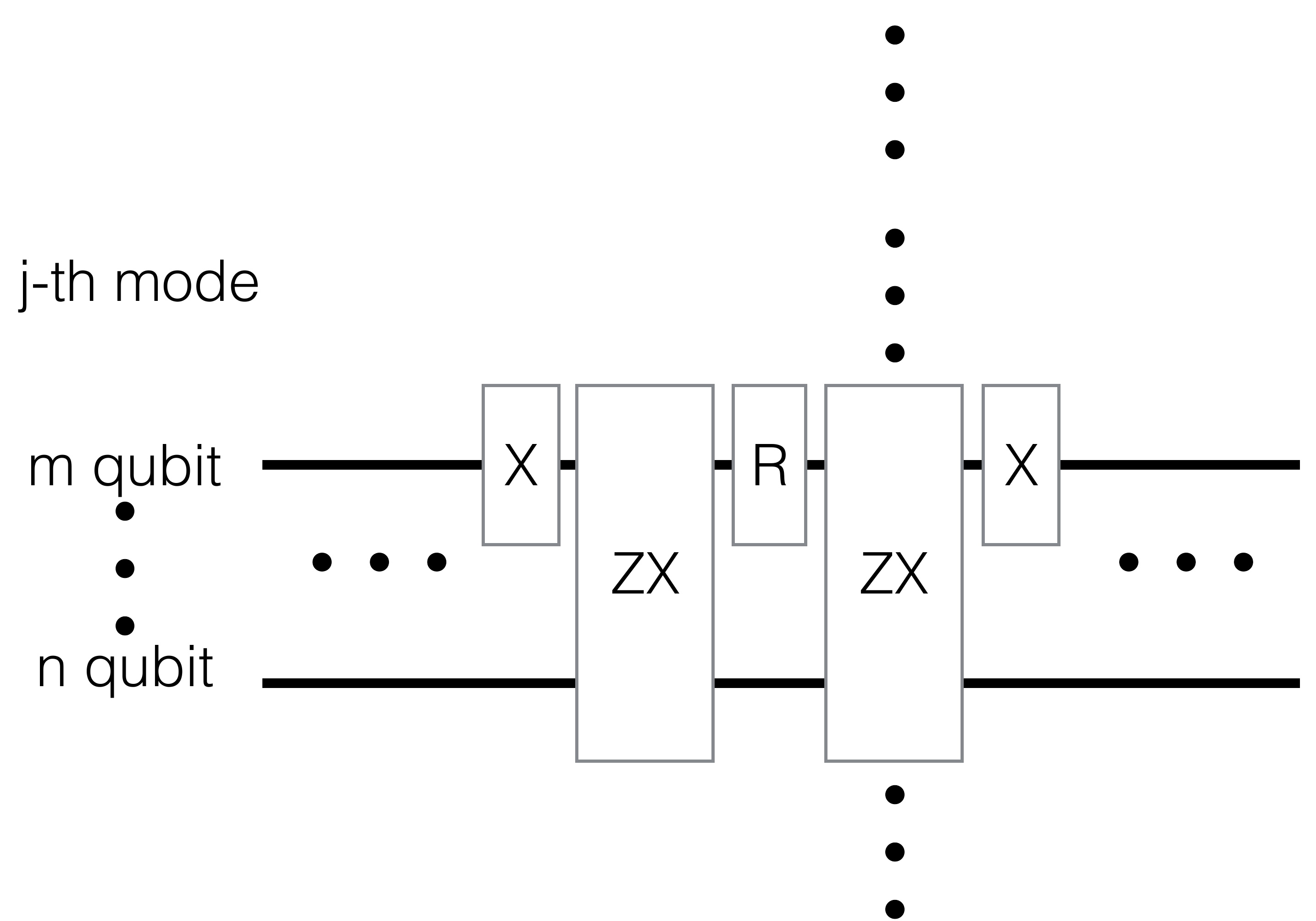} 
\caption{Sequence of gates that implement the $\sigma_z - \sigma_z$ term in Equation (\ref{eq:sz-sz}) between two qubits $m$ and $n$ out of the five qubits corresponding to the $j^{\text{th}}$ mode. $R=e^{-i\frac{\chi_j n m t_s}{2}\sigma_z^{m,j}}$, $ZX=e^{i\frac{\pi}{4}\sigma_z^{(n,j)}\sigma_x^{(m,j)}}$, and $X=e^{i\frac{\pi}{4}\sigma_x^{(m,j)}}$. This sequence is the building block of the digitalization of molecular Hamiltonians.}
 \label{Fig3}
 \end{figure}

By using the gate-decomposition techniques explained above, we can write any $\sigma_z - \sigma_z$ time evolution term as:
\begin{equation}
e^{i\frac{\chi_j n m t_s}{2}\sigma_z^{n,j}\sigma_z^{m,j}}=U^\dagger e^{-i\frac{\chi_j n m t_s}{2}\sigma_z^{m,j}}U,\label{eq:sz-sz}
\end{equation}
where:
\begin{equation}
U= e^{i\frac{\pi}{4}\sigma_x^{m,j}}e^{i\frac{\pi}{4}\sigma_z^{n,j}\sigma_x^{m,j}}.
\end{equation}
See Figure \ref{Fig3} for a generic scheme of this gate decomposition.

Adding more modes to the simulation does not change the fact that all the terms commute. Each~five-qubit string will be governed by a similar Hamiltonian, but with different $\omega_j$ and, in~general,~$\chi_j$.

\section{Conclusions}

There is a growing interest in building up large networks of thousands of qubits for groundbreaking applications in quantum simulation and quantum computing. In parallel, continuous-variable quantum information processing and quantum computation has motivated a~renewed effort in the construction of large linear optical grids. In this work, we brought the latter into the realm of the former, by providing a recipe for the digitalization of bosonic linear and nonlinear optics interactions. Applications range from the dynamical Casimir effect to molecular force fields. A~boson sampler in the quantum supremacy regime would require a number of qubits that seems out of reach; however, we have shown that all the individual ingredients of a boson sampler can be digitalized, enabling the digital quantum simulation of a few-mode setup. All these phenomena, which~are typically considered in analog simulators, would then benefit in principle from the advantages of digitalization, such as error correction. Moreover, we can digitalize quantum gates that are crucial for continuous-variable quantum information processing and quantum computing. 
As an example, we have launched several experiments in IBM Q 5 Tenerife, obtaining a fidelity of 84\% for the digital quantum simulation of one building block of a beam-splitter and 58\% for the full beam-splitter.

%%%%%%%%%%%%%%%%%%%%%%%%%%%%%%%%%%%%%%%%%%

%%%%%%%%%%%%%%%%%%%%%%%%%%%%%%%%%%%%%%%%%%
%% optional
% \supplementary{The following are available online at www.mdpi.com/link, Figure S1: title, Table S1: title, Video S1: title.}
I %Please add: ``This research received no external funding'' or ``This research was funded by NAME OF FUNDER grant number XXX.'' and and ``The APC was funded by XXX''. Check carefully that the details given are accurate and use the standard spelling of funding agency names at \url{https://search.crossref.org/funding}, any errors may affect your future funding.

%%%%%%%%%%%%%%%%%%%%%%%%%%%%%%%%%%%%%%%%%%
\section*{Acknowledgments}I am indebted to Borja Peropadre and Andr\'es Agust\'i for discussions and technical help. I have received financial support through the Postdoctoral Junior Leader Fellowship Programme from ``la Caixa'' Banking Foundation (fellowship code: LCF/BQ/LR18/11640005).
 I~acknowledge use of the IBM Q Experience for this work. The views expressed are those of the author and do not reflect the official policy or position of IBM nor the IBM Q Experience team.

%%%%%%%%%%%%%%%%%%%%%%%%%%%%%%%%%%%%%%%%%%
%\authorcontributions{C. S. proposed the general idea and supervised the work and the writing of the mansucript. C.S-V. developed the ideas, made all the computations and wrote the mansucript.}

%%%%%%%%%%%%%%%%%%%%%%%%%%%%%%%%%%%%%%%%%%
%\conflictofinterests{The author declares no conflict of interest.} 

%%%%%%%%%%%%%%%%%%%%%%%%%%%%%%%%%%%%%%%%%%
%% optional
% \abbreviations{The following abbreviations are used in this manuscript:\\

% \noindent MDPI: Multidisciplinary Digital Publishing Institute\\
% DOAJ: Directory of open access journals\\
% TLA: Three letter acronym\\
% LD: linear dichroism}

%%%%%%%%%%%%%%%%%%%%%%%%%%%%%%%%%%%%%%%%%%
%% optional
%\appendix
%\section{}
%The appendix is an optional section that can contain details and data supplemental to the main text. For example, explanations of experimental details that would disrupt the flow of the main text, but nonetheless remain crucial to understanding and reproducing the research shown; figures of replicates for experiments of which representative data is shown in the main text can be added here if brief, or as Supplementary data. Mathemtaical proofs of results not central to the paper can be added as an appendix.
\section*{References} %Please relayout the reference according to the format below.
%\bibitem[Author1(year)]{ref-journal}
%Author1, T. The title of the cited article. {\em Journal Abbreviation} {\bf 2008}, {\em 10}, 142--149.
% Reference 2
%\bibitem[Author2(year)]{ref-book}
%Author2, L. The title of the cited contribution. In {\em The Book Title}; Editor1, F., Editor2, A., Eds.; Publishing House: City, Country, 2007; pp. 32--58.

%%%%%%%%%%%%%%%%%%%%%%%%%%%%%%%%%%%%%%%%%%
%\bibliographystyle{mdpi}
%
%=====================================
% References, variant A: internal bibliography
%=====================================
%\renewcommand\bibname{References}

%=====================================
% References, variant B: external bibliography
%=====================================
%\bibliography{your_external_BibTeX_file}

%%%%%%%%%%%%%%%%%%%%%%%%%%%%%%%%%%%%%%%%%%
%% optional
% \sampleavailability{Samples of the compounds ...... are available from the authors.}

%%%%%%%%%%%%%%%%%%%%%%%%%%%%%%%%%%%%%%%%%%
\end{document}